\documentclass[12pt]{article}
\usepackage{amssymb,color,graphicx}

\begin{document}

\begin{center}
{\Large
{\bfseries PHYSICAL SCIENCE}\\
A revitalization of the traditional course\\
by avatars of Hollywood in the physics classroom }

\vspace{1cm}
 Costas Efthimiou and Ralph Llewellyn \\
 Department of Physics\\
 University of Central Florida\\
 Orlando, FL 32816

\end{center}


\section{Public Attitudes and Understanding of Science}

It has been well-documented \cite{NSF2002} that most Americans
have very little understanding of science, are unable to
distinguish between science and psuedoscience, and have no clear
concept of the role of science in their daily lives. While about
90 percent if those surveyed by the National Science Foundation
(NSF) since 1979 report being interested in science and more than
80 percent (in 2001) believe that federal government should
finance scientific research, about 50 percent do not know that
Earth takes one year to go around the sun, electrons are smaller
than atoms, and early humans did not live at the same time as
dinosaurs. Statistics concerning Americans' lack of knowledge of
such things receive frequent media attention.

Answering the question ``\textit{Who is responsible?}" is not an
easy task since many factors, including social and economic
conditions, have a direct impact on the science literacy of the
public. However, scientists strongly believe that the
media---ironically, exactly those that raise concern about the
limited science literacy of the public---and, in particular, the
entertainment industry, are at least partially responsible and are
a significant source of the public's misunderstanding and faulty
knowledge of science. Unfortunately,  the public who watch
television shows and films with pseudoscientific or paraphysical
themes does not always interpret them  simply as entertainment
based in pure fiction. Due to the lack of critical thinking
skills, many people  tend to perceive the events depicted  as real
or within the reaches of science. This unchallenged manner in
which the entertainment industry portrays pseudoscientific and
paraphysical phenomena should excite great concern in the
scientific community. Not only does it amplify the public's
scientific illiteracy, but also puts at risk the public's attitude
towards science, raising the possibility that future influential
figures in our society could inadvertently cause serious damage to
mainstream physics simply through ignorance and misunderstanding.

The authors have embarked on an ambitious project to help improve
public understanding of the basic principles of physical science.
This paper reports the results of the initial phase of the
program, which was begun with several large groups of non-science
majors enrolled in the general education physical science course
at the University of Central Florida (UCF), a course with a
counterpart in nearly every college and university (and many high
schools) in the nation.

\section{On Our Course}

\subsection{The Course in Brief}
At UCF the course is 3 semester-hours and is taken by about 3000
students annually.  It has an independent 1 semester-hour
laboratory elected by about 20 percent of  those in the course. At
this stage the lab is not a part of the project.  During the
academic year the class is taught in sections of 300 to 450
students; in the summer the class sections are limited to 90
enrollees.  All sections are taught in multimedia-equipped
classrooms.

In the authors' sections, we teach the key concepts from all areas
of physics, not only the standard core of classical physics that
is presently 'physical science', but go beyond to introduce
students to the captivating discoveries of relativity, quantum
mechanics, astronomy, and cosmology. Augmenting the traditional
lectures and live demonstrations, our new ``weapon" in this effort
to improve scientific literacy is the films themselves, often the
same ones that perpetuate the incorrect understanding of science
held by the students. Using a medium that is familiar to and
universally enjoyed by the students, we employ short (5 to 8
minutes) clips from many films as the basis for discussions and
calculations for the very broad range of topics covered in the
course, including  some decidedly non-typical topics, such as time
travel, extraterrestrial civilizations, and black holes.

\subsection{The Goals of the Course}
Our goals for the course are these:
\begin{enumerate}
 \item To motivate students to think critically about science
information  presented in films.
 \item To help students learn
to distinguish between physical laws and pseudo-science.
 \item To encourage students to understand how science works and how any widely accepted
theory has been verified using the scientific method.
 \item To help students learn where the borderline  between tested and untested physics ideas
lies.
\end{enumerate}

\subsection{Other Similar Attempts}
In years past forerunners of our approach have been tried in a few
places  with varying degrees of success. The most notable case is
that of Professor L. Dubeck of Temple Universiy
\cite{Dubeck,Dubeck2}. Dubeck has used science fiction films to
teach scientific ideas to non-science majors. A real pioneer in
this approach, but well ahead of his time, his course did not
attract the attention it deserved.  We only discovered it while in
the process of developing our course for the Summer of 2002. His
approach is qualitatively; on the contrary, our approach is mostly
quantitative, although qualitative arguments are often given for
various scenes. Moreover, we have made use of a broad array of
genres,  not just science fiction films.

Other efforts have focused at the high school level \cite{Rogers,
Dennis,Dennis2}. In particular, Dennis has recently advertised his
success in an article published in Physics Teacher \cite{Dennis}
about teaching mechanics with the aid of films. Motivated by this
success, Dennis has published a book\footnote{We are grateful to
C. Dennis for sending us review copies of his book.}
\cite{Dennis2} with the hope of helping other teachers use films
in their courses, too. Our approach carries many similarities in
philosophy with that of Dennis; however, our course, taught at a
college level is more advanced and covers by far more topics than
mechanics.

A serious challenge in each case  has been to keep the course
up-to-date---for example, this seems to be the most serious
problem with the books \cite{Dubeck,Dubeck2} as expressed by the
students. This challenge we think our approach and careful
planning will meet.

\section{On Our Use of Films}

The course syllabus outlines the topics to be covered during the
term and also includes a schedule of films to be viewed as
homework by the students.
\begin{figure}[h!]
\begin{center}
\includegraphics[width=3cm]{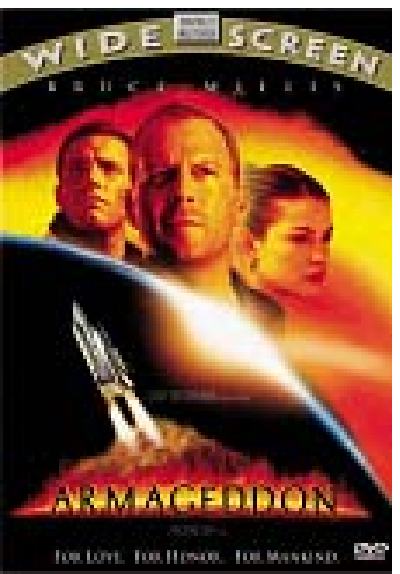}
\includegraphics[width=3cm]{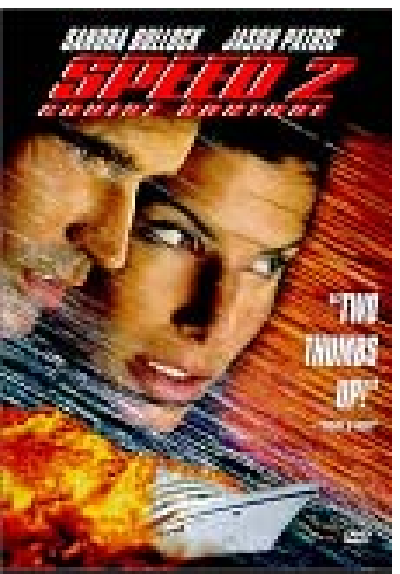}
\includegraphics[width=3cm]{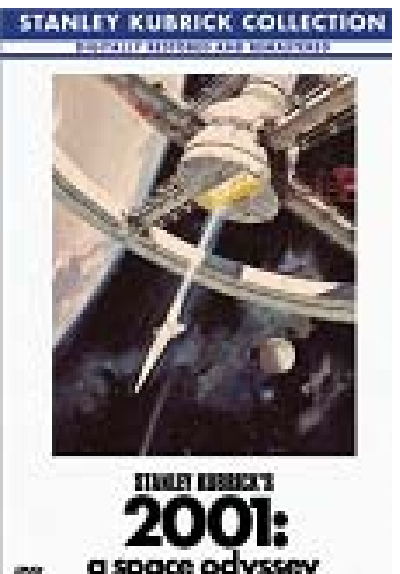}\\[1mm]
\includegraphics[width=3cm]{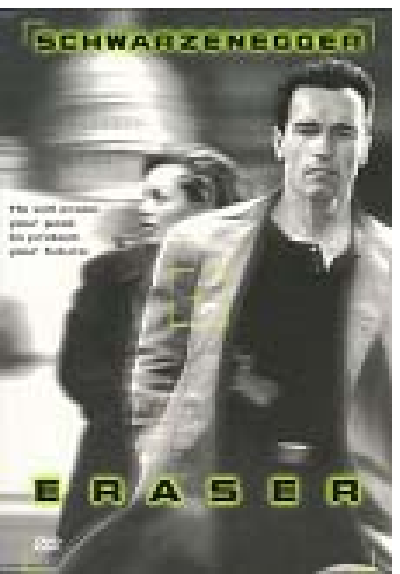}
\includegraphics[width=3cm]{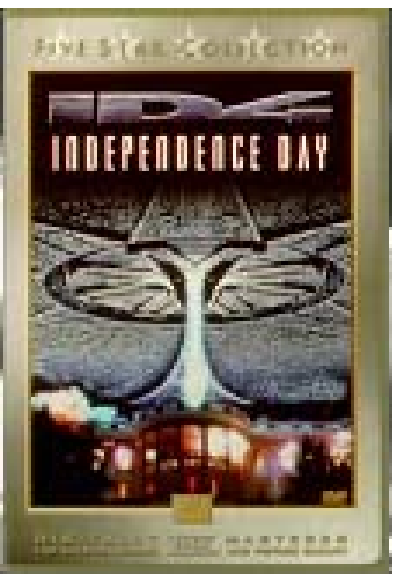}
\includegraphics[width=3cm]{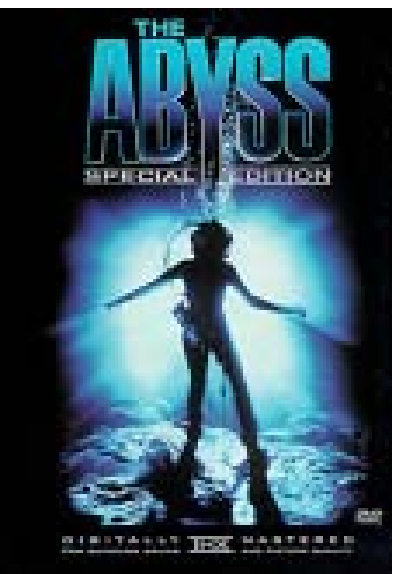}\\[1mm]
\includegraphics[width=3cm]{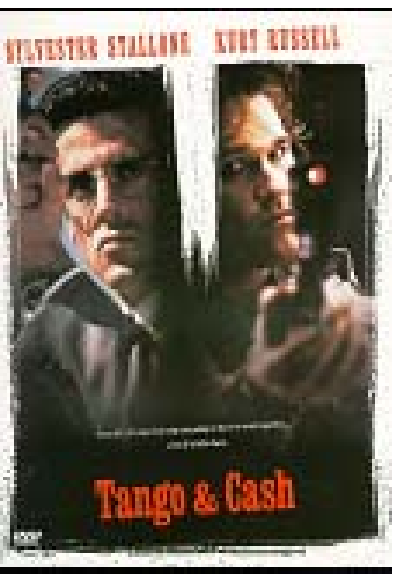}
\includegraphics[width=3cm]{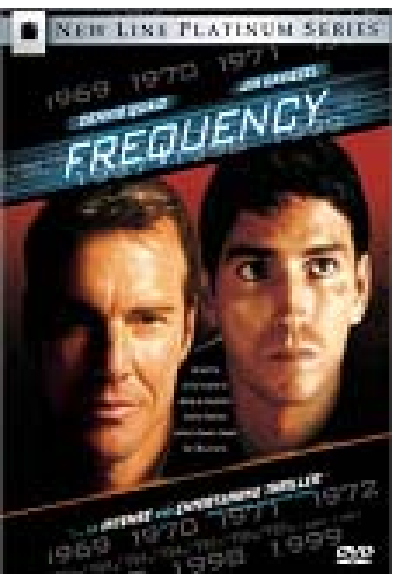}
\includegraphics[width=3cm]{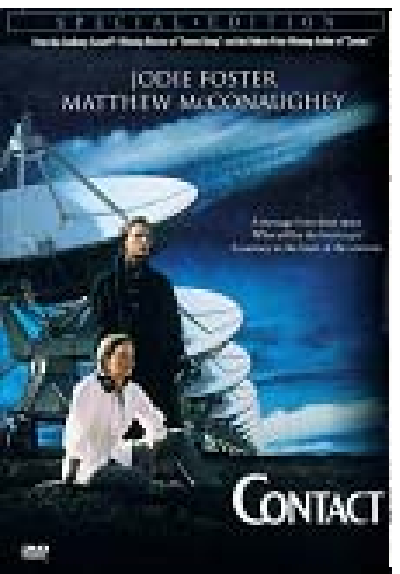}
\end{center}
\caption{The main nine films that were used in the initial phase
         of the course at
         UCF. Clips from several other Hollywood movies are shown to
         students, as well as clips from scientific documentaries
         and IMAX films. However, the additional films are not required to be
         pre-viewed at home.}
\end{figure}
Each film contains scenes concerned with one or more (usually
more) of the topics. For example, listed below are the films and
topics used during the initial phase of the project:
 \begin{enumerate}
 \item \textsf{Speed 2} \cite{Speed2}: speed and acceleration, collisions.
 \item \textsf{Armageddon} \cite{Armageddon}: momentum, energy, comets and
       asteroids.
 \item \textsf{Eraser} \cite{Eraser}: momentum, free fall, electromagnetic
       radiation.
 \item \textsf{2001: A Space Odyssey} \cite{2001}: centripetal and centrifugal
       force, artificial gravity, zero gravity.
 \item \textsf{The Abyss} \cite{Abyss}: hydrostatic and atmospheric pressure,
       effects of pressure on humans and objects.
 \item \textsf{Independence Day} \cite{ID}: potential energy, conservation
       of energy, pressure, tidal forces.
 \item \textsf{Tango and Cash} \cite{T+C}: momentum, electricity, physiological effects
       on humans.
 \item \textsf{Frequency} \cite{Frequency}: magnetism, motion of particles in
       magnetic fields, Aurora Borealis, solar wind, wormholes,
       time travel.
 \item \textsf{Contact} \cite{Contact}: relativity, space travel, wormholes, life beyond
       Earth, Drake's formula.
 \end{enumerate}
The films assigned, about one per week, are popular ones available
on DVDs or videotapes from stores or, in many cases,
libraries\footnote{The UCF library currently owns 10 DVDs from
each movie used in the course.}. Films are viewed at home as
homework prior to discussion of the related topics in class; as
mentioned, only short clips are used in class. We found that
viewing the homework films quickly became occasions for student
group or family pizza parties!  This helped alleviate a potential
problem of having enough copies of each film locally available.
Sufficient available copies turned out not to be a problem as the
numerous video stores tend to have many copies of the films we
use.

In class integrated with the lecture, PowerPoint slides and
demonstrations, we focus on analyzing the topic of the day using
clips from films that have examples of  that principle. For
example, students might watch \textsf{The Abyss}.  Then in class,
using a couple of clips from the film to ``set the scene", we
discuss pressure in a fluid as a function of depth, how deep an
object (e.g., a person or submarine) can descend, and the
propagation of sound in water. Following that discussion, a
physics single-concept film may be shown to illustrate what the
movie got right and what it got wrong.  After class, the
instructor's analysis of each clip is placed on the course web
site so that students may review the discussion and calculations.
An additional advantage of this approach is that the films we use
can be varied each term to incorporate new releases and to focus
attention on particular concepts. Below is an example of how we
use particular clips from \textsf{Armageddon} in discussions of
asteroids, momentum conservation and energy.

\vspace{5mm}
\hrule \hrule

\vspace{1mm}

\small
 \noindent {\bfseries EXAMPLE}: In \textsf{Armageddon}
\cite{Armageddon} an asteroid of the size of Texas is on a
collision course with Earth. In \textit{Chapter 7: Training
Begins} of \textsf{Armageddon}, the flight plan of NASA and the
method for the destruction of the asteroid are described (see
segment from 43:04 to 45:34): the team must land on the asteroid,
drill a hole, plant a nuclear bomb, and then detonate it before
the `zero barrier'. If the  detonation happens after  the `zero
barrier', then the asteroid fragments will collide with Earth.
 This segment is discussed twice in class, once when the
concepts of momentum are presented (paragraph 1 below), and once
when the concepts of energy are presented (paragraphs 3 and 4
below). In the second case, the calculation ends with an
evaluation on the feasibility of the plan. Prior to the second
time, another quick clip (\textit{Chapter 4: NYC Hit Hard},
segment from 10:20 to 11:40) is shown in class which motivates a
discussion on asteroids and the calculation of the mass of the
asteroid seen in the film (paragraph 2 below). These discussions
are repeated here in order to demonstrate our methods.

\begin{figure}[h!]
\includegraphics[width=5cm]{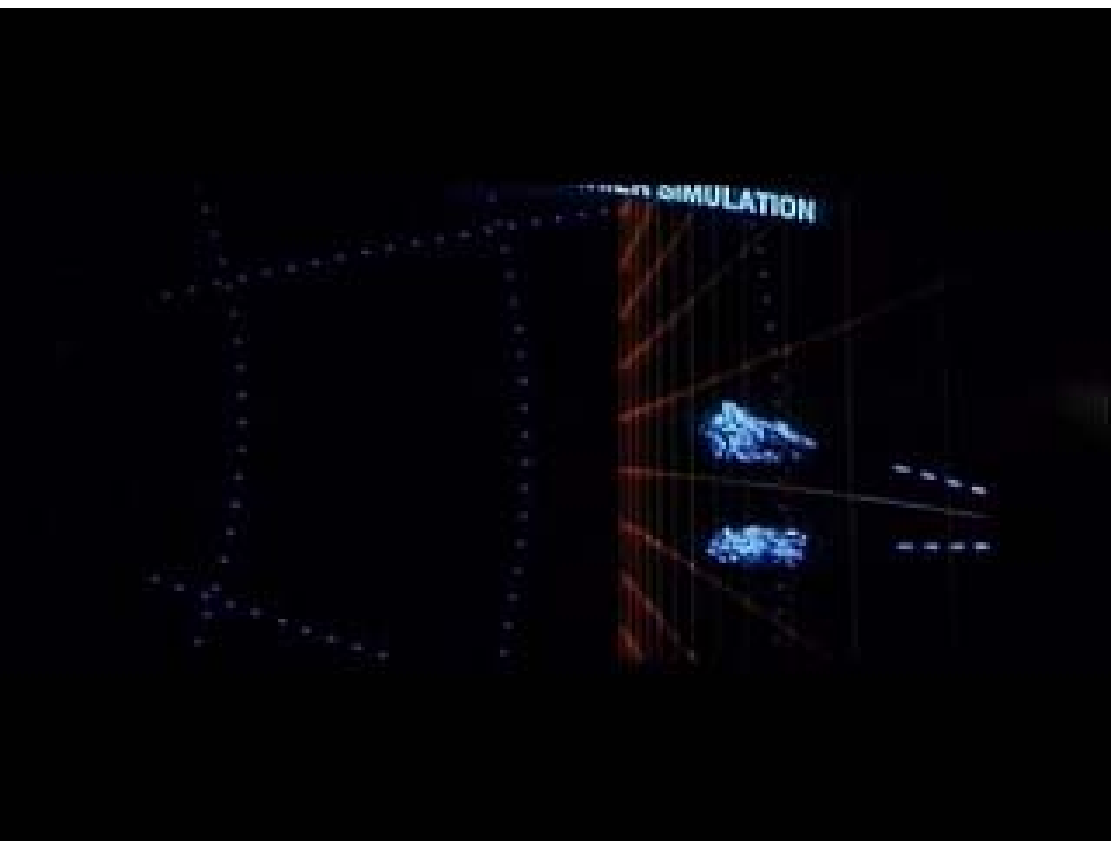}
\includegraphics[width=5cm]{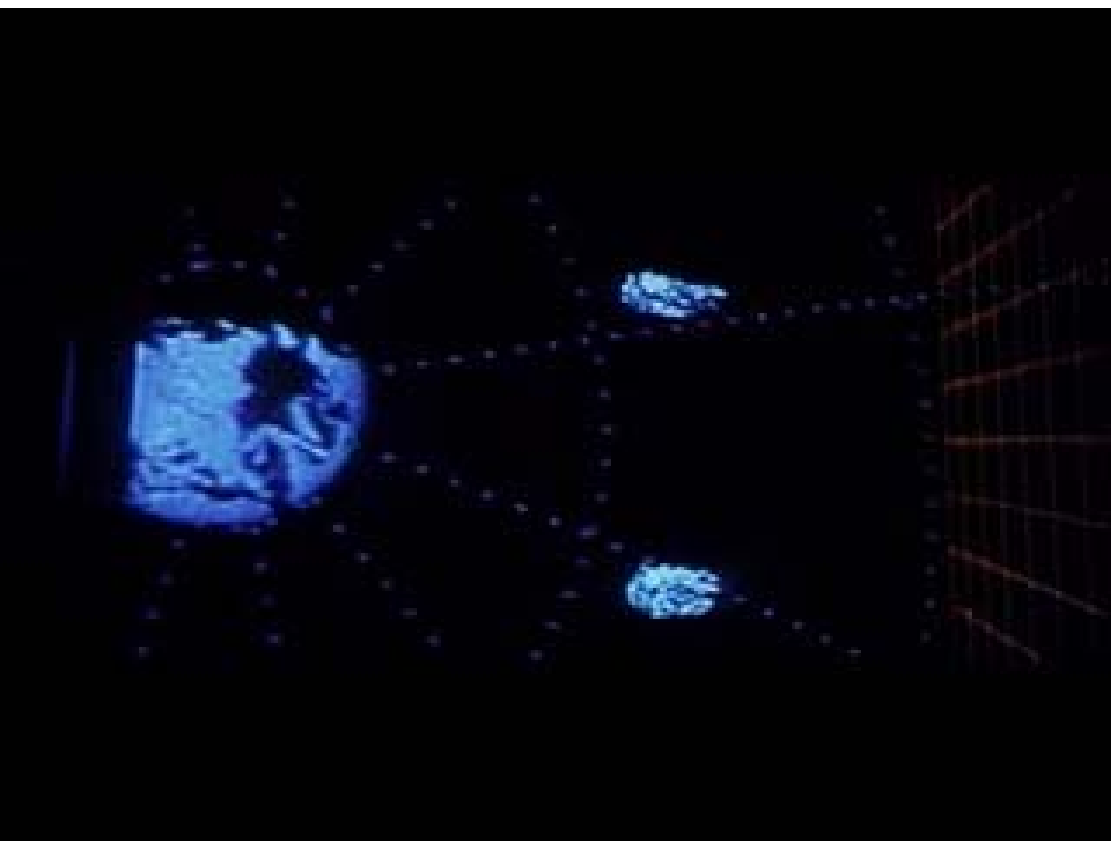}
\includegraphics[width=5cm]{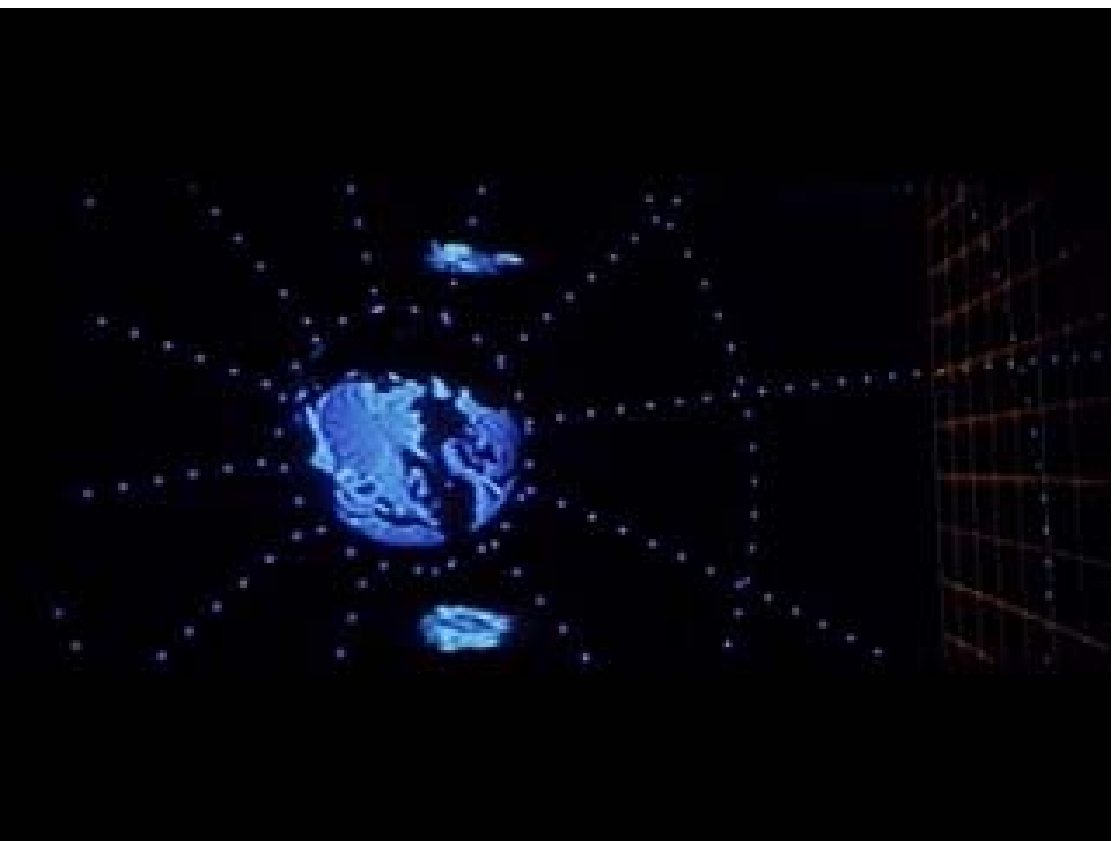}
\caption{\textsf{Armageddon} NASA simulations show that the
successful detonation of the bomb before the zero barrier will
result in splitting the asteroid in two fragments that will be
deflected away from Earth.}
\end{figure}

\begin{figure}[h!]
\begin{center}
\includegraphics[width=5cm]{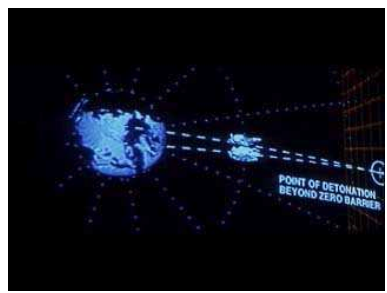}
\end{center}
\caption{NASA simulations show that  detonation of the bomb after
the zero barrier will result in splitting the asteroid in two
fragments that will not be deflected enough to avoid collision
with Earth.}
\end{figure}

\S1. {\bfseries Momentum Conservation}: Following the NASA
simulation and to simplify the analysis, we assume that the
asteroid splits in exactly two fragments. Moreover, we assume that
the detonation of the nuclear bomb will force the two fragments to
deflect perpendicular to the direction of the original  motion at
speeds that depend on the size of the bomb. (We shall indicate the
direction of motion as $x$-direction and the perpendicular
direction as $y$-direction.) The two fragments will still continue
to move towards Earth with the $22,500~mi/h$ speed of the
asteroid. This is guaranteed by momentum conservation. The
relation of the fragment speeds in the $y$-direction is also
determined by momentum conservation. Assuming that the two
fragments have equal mass, they will acquire equal speeds in the
$y$-direction, such that the total momentum along this direction
maintains the value zero that it had before the
detonation.\\[1mm]

\S2. {\bfseries Mass of the asteroid}: For a quantitative
calculation on the \textsf{Armageddon} NASA plan, we shall need
the mass of the asteroid. In the movie we are told that the
asteroid is of the size of Texas. The area of Texas is
$691,027km^2$ \cite{Grolier}. Assuming that Texas is a square,
this would give a length of $831.3km$  for the side of the square.
This gives an average length. In reality, Texas is stretching
$1,244Km$ from east to west and $1,289km$ from north to south
\cite{Grolier}. To be on the conservative side, we will adopt the
average size $831.3km$ for our calculations. Even so, only one
known asteroid is of such size; the largest asteroid, Ceres, has a
diameter of about $940km$. In fact only two dozen  asteroids have
diameters $200km$ or greater. Even if we assemble all known
asteroids to  a unique big asteroid, its diameter will be no more
than $1,500km$ and its mass less than 1/10 that of Earth's moon
\cite{Chaisson}. However, to give the director the benefit of
doubt, we shall accept the existence of an asteroid of the size
given on collision course with Earth.

If the asteroid is a cube, its volume is $5.7 \times 10^8 km^3$.
If it is a sphere, then its volume is $18 \times 10^8 km^3$. Large
asteroids have spherical shape; smaller asteroids have irregular
shapes \cite{Chaisson}. However, let's assume that the volume of
the asteroid in the film is somewhere in between the two values
given above, say $10 \times 10^8 km^3 = 10^{18} m^3$.

Overall, about 15 percent of all asteroids are silicate (i.e.
composed of rocky material), 75 percent are carbonaceous (i.e.
contain carbon), and 10 percent are other types (such as those
containing large fractions of iron). Stony asteroids look very
much like terrestrial rocks. So, we might assume that the  density
of the asteroid is approximately equal to that of Earth:
$5500kg/m^3$. However, we know from observations that asteroid Ida
has a density of $2200$--$2900kg/m^3$ and asteroid Mathilde has a
density of about $1400kg/m^3$. This especially low density of
Mathilde might be due to a porous interior \cite{Chaisson}. To be
conservative, we shall assume a density of $2000kg/m^3$ for the
film asteroid. In other words, ever cubic meter of the asteroid
would have a mass of $2000kg$. The total mass of the movie
asteroid would be $2000\times 10^{18} kg$ or $2\times 10^{21} kg$.
We have assumed that, after the detonation of the nuclear bomb,
the asteroid breaks in two pieces which have equal mass. This
means that each
piece would have a mass of $10^{21} kg$.\\[1mm]

\S3. {\bfseries Energy of nuclear weapons}:In the film a nuclear
bomb is used to split the asteroid. The explosive TNT has become
the standard means for showing the disastrous power of  nuclear
weapons. In particular, one ton (t) of TNT releases
$$
  4.2 \times 10^9~Joules
$$
of energy. Yet, one ton  is a small quantity of TNT to describe
modern bombs; kilotons (kt) or megatons (Mt) of TNT are units more
appropriate:
$$
    1kt=1000\,t~,~~~1Mt=1,000,000\,t~.
$$
The Hiroshima bomb was equivalent to $12kt$ of TNT or
$$
  5 \times 10^{13}~Joules.
$$
A modern nuclear bomb is equivalent to $20Mt$ of TNT. This is
equal to 1667 Hiroshima bombs and releases about
$$
  8.4 \times 10^{16}~Joules.
$$
Today the nations which have such weapons no longer spend effort
to increase the destructive power of nuclear bombs as it is
already immense.
Instead, they focus on reducing the size of the weapons.\\[1mm]

\S4. {\bfseries Deflection of the fragments}: Returning to the
movie, let's assume that the nuclear bomb that was carried to the
asteroid was equivalent to 100,000 Hiroshima bombs. For the
reasons we have already explained, this is a really generous
assumption in favor of the director. As always, we want to be nice
to him. Upon detonation, the energy released will be
$$
  5 \times 10^{18}~Joules.
$$
Part of this energy will be used to break the asteroid into two
pieces. However, once more, we will be generous to the director
and we will ignore this fact, even though the energy needed to do
so is significant. Therefore, we will assume that all the energy
becomes kinetic energy of the two fragments; each fragment will be
given an amount of
$$
  2.5 \times 10^{18}~Joules.
$$
Moreover, as  stated above, we  assume that all energy becomes
kinetic energy associated with motion in the $y$-direction and no
amount is spent to push the fragments in the $x$-direction.

Knowing the mass of each fragment and the kinetic energy of its
motion in the $y$-direction, we can find the speed of the fragment
in this direction:
$$
   KE = {1\over2}\, m \, v^2
   ~\Rightarrow~
  v= \sqrt{2\,KE\over m}~.
$$
If we substitute the numbers and do the calculation we find that
$$
    v = 0.07~{m\over s}~.
$$
Therefore, each fragment moves in the $y$-direction ${7\over100}m$
(or $7cm$) every second. In two hours or $7200s$, the time to
reach Earth, each fragment will move
$$
   {7\over100}~{m\over s}\times7200s = 504~m.
$$
Now compare this with the radius of the Earth:
$$
   6,500km = 6,500,000m
$$
which is how far each fragment must move  in  the $y$-direction in
order to just miss the solid Earth (but it would still go through
the atmosphere and release a lot of energy).

So we see that applying momentum conservation and reasonable (and
generous!) assumptions about the size of the nuclear bomb used,
there is no way that this plan would save Earth. The two fragments
of the asteroid would be barely few city blocks apart when they
collided with Earth. Such proposals appear in the press from time
to time, but now the students learnt how to critically analyze
such suggestions made by well-meaning, but scientifically,
illiterate contemporaries.

The above discussion, besides the explicitly made assumptions,
ignores the gravitational attraction of the two fragments that
will decrease further the deflection, the tidal effects on Earth,
and the perturbation introduced by the asteroid in the motion of
the Earth-moon system. These effects are discussed further in
class, partly qualitatively and partly quantitatively.

\normalsize

\vspace{1mm} \hrule \hrule

\vspace{5mm}

As is obvious from our example, the quantitative analyses we use
are often more sophisticated than what is expected from Physical
Science students. However, we have discovered that, since the
analysis is strongly correlated to the film, no complaints are
generated. A similar attempt to solve such a problem in the
traditional course would generate many unpleasant feelings and
would make the course very unpopular.

Besides the films required to be previewed by the students at
home, the instructors make use of clips from various other movies,
IMAX films, and scientific documentaries. For example, a clip from
\textsf{Deep Impact} is used to demonstrate the tsunamis after a
collision of a comet with Earth, clips from the IMAX film
\textsf{Mission to Mir} \cite{Mir} is shown in order to illustrate
life under ``zero gravity" conditions, and clips from the PBS
documentary \textsf{Life Beyond Earth} \cite{LBE} are shown to
discuss our quest for life in the Universe.

\section{Student Response and Performance Results}

So, how do the students perceive the course? The instructors  made
use  in class of an electronic  personal response system whereby
each student could respond immediately to questions posted by the
instructors, their responses being automatically recorded and
tabulated by an in-class computer. This system, besides its
pedagogical value, was used to obtain data on the student's
feelings and reactions on the course and to record attendance. The
results for several course-evaluation questions are shown in
Tables \ref{table:data1} through \ref{table:data4}.

\begin{table}[h!]
\begin{center}
\begin{tabular}{|c|c|c|c|}\hline
 \multicolumn{4}{|c|}{SUMMER 2002} \\ \hline
 strongly agree & agree & disagree & strongly disagree \\ \hline
 38.16\% & 52.63\% & 5.26\% & 3.95\% \\ \hline
\end{tabular}

\vspace{5mm}

\begin{tabular}{|c|c|c|c|c|}\hline
 \multicolumn{5}{|c|}{FALL 2002} \\ \hline
 strongly agree & agree & no opinion & disagree & strongly disagree \\ \hline
 28.06\% & 48.98\% & 13.27\% & 3.57\% & 6.12\% \\ \hline
\end{tabular}
\end{center}

\caption{Data on the question ``The films were well-chosen to
         include a broad range of science ideas".}
\label{table:data1}
\end{table}

\begin{table}[h!]
\begin{center}
\begin{tabular}{|c|c|c|c|}\hline
 \multicolumn{4}{|c|}{SUMMER 2002} \\ \hline
 strongly agree & agree & disagree & strongly disagree \\ \hline
 28.95\% &  60.52\% & 9.21\% & 1.32\% \\ \hline
\end{tabular}

\vspace{5mm}

\begin{tabular}{|c|c|c|c|c|}\hline
 \multicolumn{5}{|c|}{FALL 2002} \\ \hline
 strongly agree & agree & no opinion & disagree & strongly disagree \\ \hline
 19.72\% & 53.99\% & 17.37\% & 6.10\% & 2.82\% \\ \hline
\end{tabular}
\end{center}

\caption{Data on the question ``The topics selected from the
         movies for physics analysis were interesting".}
\end{table}

\begin{table}[h!]
\begin{center}
\begin{tabular}{|c|c|c|c|}\hline
 \multicolumn{4}{|c|}{SUMMER 2002} \\ \hline
 strongly agree & agree & disagree & strongly disagree \\ \hline
 77.92\% & 10.39\% & 9.09\% & 2.60\% \\ \hline
\end{tabular}

\vspace{5mm}

\begin{tabular}{|c|c|c|c|c|}\hline
 \multicolumn{5}{|c|}{FALL 2002} \\ \hline
 strongly agree & agree & no opinion & disagree & strongly disagree \\ \hline
 56.88\% & 26.61\% & 6.88\% & 4.13\% & 5.50\% \\ \hline
\end{tabular}
\end{center}

\caption{Data on the question ``The instructors should develop
         this course further since it is more interesting than the standard
         physical science course".}
\end{table}

\begin{table}[h!]
\begin{center}
\begin{tabular}{|c|c|c|c|}\hline
 \multicolumn{4}{|c|}{SUMMER 2002} \\ \hline
 strongly agree & agree & disagree & strongly disagree \\ \hline
 66.24\% & 27.27\% & 5.19\% & 1.30\% \\ \hline
\end{tabular}

\vspace{5mm}

\begin{tabular}{|c|c|c|c|c|}\hline
 \multicolumn{5}{|c|}{FALL 2002} \\ \hline
 strongly agree & agree & no opinion & disagree & strongly disagree \\ \hline
 48.64\% & 30.91\% & 7.72\% & 4.55\% & 8.18\% \\ \hline
\end{tabular}
\end{center}

\caption{Data on the question ``I would recommend to my friends
         that they take this course".}
\label{table:data4}
\end{table}

Besides the data collected by the personal response system,
students have expressed strong support through the standard
end-of-term course evaluations and through ---unsolicited---
comments in their term papers.

Of course, even if the students embrace a new idea
enthusiastically, it does not mean that their performance will be
better than their performance in the traditional course (where the
majority of them really struggle). The effect needs to documented.
In Table \ref{table:performance}, we list the performance of UCF
Physical Science students in two almost identical classes, one
taught in the traditional way and one taught using movies. The
important parameters in both classes are identical: the classes
have the same size, they were taught by the same instructor (C.E.)
who used  similar PowerPoint lectures, same demonstrations, gave
similar exams. The classes were even taught in the same
auditorium. The classes also covered similar material with the
same textbook \cite{Bolemon} required in the traditional class and
recommended in the Physics in Films class. The Physics in Films
class used in addition a supplement \cite{Dubeck}. The material of
the Physics in Films class was almost identical. There were some
differences in order to make the content of the course more
exciting: a few topics of the traditional syllabus were omitted
and they were substituted by topics that captivate the imagination
of the students. Topics that were covered in the traditional
course but omitted in the Physics in Films course were the
extensive treatment of heat and temperature and extensive
discussion of elements and the periodic table. Instead, topics
from astronomy (comets and asteroids, life and intelligent life
beyond Earth), as well as some topics from modern science were
added (elements of special and general relativity, space and time
travel).

\begin{table}[h!]
\begin{center}
\begin{tabular}{|c|c|c|c|c|}\hline
 \multicolumn{5}{|c|}{FALL 2001: traditional Physical Science, 295 students} \\ \hline
                     & TEST 1 & TEST 2 & TEST 3 & FINAL \\ \hline
  average            & 49.34  & 65.33  & 58.18  & 59.44 \\ \hline
  standard deviation & 13.22  & 16.09  & 15.88  & 11.67 \\ \hline
\end{tabular}

\vspace{5mm}

\begin{tabular}{|c|c|c|c|c|}\hline
 \multicolumn{5}{|c|}{FALL 2002: Physics in Films, 292 students} \\ \hline
                     & TEST 1 & TEST 2 & TEST 3 & FINAL \\ \hline
  average            & 74.92  & 67.68  & 75.68  & 72.82 \\ \hline
  standard deviation & 14.36  & 16.92  & 14.08  & 12.84 \\ \hline
\end{tabular}
\end{center}

\caption{Data on similar exams from two Physical Science classes
         of identical size taught by the same instructor.
         The material of the two classes was almost identical.
         (See discussion in article.)
         Exams are normalized to a maximum of 100 points. The differences
         are both dramatic and significant.}
\label{table:performance}
\end{table}

\section{On the Drawing Board}

The authors are currently writing a physical science textbook that
incorporates the 'physics in films' concept. It will be complete
with an instructors' CD containing analysis notes for every scene
used in the book to discuss concepts in physical science and a
tutorial on making your own analysis of scenes from favorite or
new movies. In addition to physics, this approach to teaching
basic concepts at the non-major introductory level can be applied
in many other disciplines.  Table \ref{table:1} lists several
possibilities with examples of films containing pertinent scenes.

\begin{table}[h!]
 \begin{center}
 \begin{tabular}{|l|l|}\hline
  {\bfseries Discipline} & {\bfseries Films} \\ \hline
  Astronomy & Armageddon, Contact, Deep Impact, ... \\ \hline
  Biology   & Planet of the Apes, Spiderman, ... \\ \hline
  Mathematics & Contact, Pay It Forward, Pi, A Beautiful Mind, ... \\ \hline
 \end{tabular}
 \end{center}
 \caption{Use of films in other disciplines. We have listed three other areas besides Physics.
          However, the possibilities are really unlimited:
          Archaeology, Chemistry, Computer Science, Engineering, Forensic Science, History, Law,
          Philosophy, etc.}
 \label{table:1}
\end{table}


\begin{thebibliography}{10}
\bibitem{NSF2002}
   \textit{Science \& Engineering Indicators 2002},
   \texttt{http://www.nsf.gov/sbe/srs/seind02}.
\bibitem{Dubeck}
    \textsc{L.W. Dubeck, S.E. Moshier, J.E. Boss},
    \textsf{Fantastic Voyages: learning science through science
    fictions films}, Springer 1994.
\bibitem{Dubeck2}
    \textsc{L.W. Dubeck, S.E. Moshier, J.E. Boss},
    \textsf{Science in Cinema: teaching science fact through science
    fictions films}, Teachers College 1988.
\bibitem{Speed2}
    \textsf{Speed2}, Widescreen DVD, 20th Century Fox 1998.
\bibitem{Armageddon}
    \textsf{Armageddon}, Widescreen DVD, Touchstone Pictures 1997.
\bibitem{Eraser}
    \textsf{Eraser}, DVD, Warner Bros. 1996.
\bibitem{2001}
    \textsf{2001: A Space Odyssey}, Widescreen DVD, Warner Bros. 2001.
\bibitem{Abyss}
    \textsf{The Abyss}, Special Edition DVD, 20th Century Fox 2002.
\bibitem{ID}
    \textsf{Independence Day}, Widescreen DVD, 20th Century Fox 2002.
\bibitem{T+C}
    \textsf{Tango \& Cash}, DVD, Warner Bros. 1997.
\bibitem{Frequency}
    \textsf{Frequency}, Platinum Series DVD, New Line 2001.
\bibitem{Contact}
    \textsf{Contact}, Widescreen DVD, Warner Bros. 1997.
\bibitem{Mir}
    \textsf{Mission to Mir}, IMAX DVD, Warner Bros. 1997.
\bibitem{LBE}
    \textsf{Life Beyond Earth}, PBS DVD, Warner Bros. 2000.
\bibitem{DI}
    \textsf{Deep Impact}, Widescreen DVD, Paramount 1998.
\bibitem{Rogers}
    \textsc{T. Rogers}, personal communication.
\bibitem{Dennis}
    \textsc{C.M. Dennis, Jr}, \textit{Start Using ``Hollywood
    Physics" in your Classroom}, Physics Teacher {\bfseries 40}
    (2002) 420.
\bibitem{Dennis2}
    \textsc{C.M. Dennis, Jr}, \textsf{Hollywood Physics:
    Mechanics}, Fidget Publications 2001.

\bibitem{Grolier}
  \textsf{Grolier}, Encyclopedia of Knowledge.
\bibitem{Chaisson}
   \textsc{E. Chaisson, S. McMillan}, \textsf{Astrnomy: A Beginner's Guide to the
   Universe}, Pentice Hall 2001.
\bibitem{Bolemon}
  \textsc{J. Bolemon}, \textsf{The Physics Around You}, McGraw-Hill
  2001.
\end{thebibliography}
\end{document}